\documentclass[12pt]{article}
\usepackage{epsfig}
\usepackage{amsmath}
\usepackage{hhline}
\usepackage{amssymb}
\usepackage{times}
\usepackage{cite}

\newlength{\dinwidth}
\newlength{\dinmargin}
\begin{document}  
\newcommand{\pom}{{I\!\!P}}
\newcommand{\reg}{{I\!\!R}}
\newcommand{\slowpi}{\pi_{\mathit{slow}}}
\newcommand{\fiidiii}{F_2^{D(3)}}
\newcommand{\fiidiiiarg}{\fiidiii\,(\beta,\,Q^2,\,x)}
\newcommand{\n}{1.19\pm 0.06 (stat.) \pm0.07 (syst.)}
\newcommand{\nz}{1.30\pm 0.08 (stat.)^{+0.08}_{-0.14} (syst.)}
\newcommand{\fiidiiiful}{F_2^{D(4)}\,(\beta,\,Q^2,\,x,\,t)}
\newcommand{\fiipom}{\tilde F_2^D}
\newcommand{\ALPHA}{1.10\pm0.03 (stat.) \pm0.04 (syst.)}
\newcommand{\ALPHAZ}{1.15\pm0.04 (stat.)^{+0.04}_{-0.07} (syst.)}
\newcommand{\fiipomarg}{\fiipom\,(\beta,\,Q^2)}
\newcommand{\pomflux}{f_{\pom / p}}
\newcommand{\nxpom}{1.19\pm 0.06 (stat.) \pm0.07 (syst.)}
\newcommand {\gapprox}
   {\raisebox{-0.7ex}{$\stackrel {\textstyle>}{\sim}$}}
\newcommand {\lapprox}
   {\raisebox{-0.7ex}{$\stackrel {\textstyle<}{\sim}$}}
\def\gsim{\,\lower.25ex\hbox{$\scriptstyle\sim$}\kern-1.30ex%
\raise 0.55ex\hbox{$\scriptstyle >$}\,}
\def\lsim{\,\lower.25ex\hbox{$\scriptstyle\sim$}\kern-1.30ex%
\raise 0.55ex\hbox{$\scriptstyle <$}\,}
\newcommand{\pomfluxarg}{f_{\pom / p}\,(x_\pom)}
\newcommand{\dsf}{\mbox{$F_2^{D(3)}$}}
\newcommand{\dsfva}{\mbox{$F_2^{D(3)}(\beta,Q^2,x_{I\!\!P})$}}
\newcommand{\dsfvb}{\mbox{$F_2^{D(3)}(\beta,Q^2,x)$}}
\newcommand{\dsfpom}{$F_2^{I\!\!P}$}
\newcommand{\gap}{\stackrel{>}{\sim}}
\newcommand{\lap}{\stackrel{<}{\sim}}
\newcommand{\fem}{$F_2^{em}$}
\newcommand{\tsnmp}{$\tilde{\sigma}_{NC}(e^{\mp})$}
\newcommand{\tsnm}{$\tilde{\sigma}_{NC}(e^-)$}
\newcommand{\tsnp}{$\tilde{\sigma}_{NC}(e^+)$}
\newcommand{\st}{$\star$}
\newcommand{\sst}{$\star \star$}
\newcommand{\ssst}{$\star \star \star$}
\newcommand{\sssst}{$\star \star \star \star$}
\newcommand{\tw}{\theta_W}
\newcommand{\sw}{\sin{\theta_W}}
\newcommand{\cw}{\cos{\theta_W}}
\newcommand{\sww}{\sin^2{\theta_W}}
\newcommand{\cww}{\cos^2{\theta_W}}
\newcommand{\trm}{m_{\perp}}
\newcommand{\trp}{p_{\perp}}
\newcommand{\trmm}{m_{\perp}^2}
\newcommand{\trpp}{p_{\perp}^2}
\newcommand{\alp}{\alpha_s}

\newcommand{\alps}{\alpha_s}
\newcommand{\sqrts}{$\sqrt{s}$}
\newcommand{\LO}{$O(\alpha_s^0)$}
\newcommand{\Oa}{$O(\alpha_s)$}
\newcommand{\Oaa}{$O(\alpha_s^2)$}
\newcommand{\PT}{p_{\perp}}
\newcommand{\JPSI}{J/\psi}
\newcommand{\sh}{\hat{s}}
\newcommand{\uh}{\hat{u}}
\newcommand{\MP}{m_{J/\psi}}
\newcommand{\PO}{I\!\!P}
\newcommand{\xbj}{x}
\newcommand{\xpom}{x_{\PO}}
\newcommand{\ttbs}{\char'134}
\newcommand{\xpomlo}{3\times10^{-4}}  
\newcommand{\xpomup}{0.05}  
\newcommand{\dgr}{^\circ}
\newcommand{\pbarnt}{\,\mbox{{\rm pb$^{-1}$}}}
\newcommand{\gev}{\,\mbox{GeV}}
\newcommand{\WBoson}{\mbox{$W$}}
\newcommand{\fbarn}{\,\mbox{{\rm fb}}}
\newcommand{\fbarnt}{\,\mbox{{\rm fb$^{-1}$}}}
%
%
\newcommand{\qsq}{\ensuremath{Q^2} }
\newcommand{\gevsq}{\ensuremath{\mathrm{GeV}^2} }
\newcommand{\et}{\ensuremath{E_t^*} }
\newcommand{\rap}{\ensuremath{\eta^*} }
\newcommand{\gp}{\ensuremath{\gamma^*}p }
\newcommand{\dsiget}{\ensuremath{{\rm d}\sigma_{ep}/{\rm d}E_t^*} }
\newcommand{\dsigrap}{\ensuremath{{\rm d}\sigma_{ep}/{\rm d}\eta^*} }
\def\Journal#1#2#3#4{{#1} {\bf #2} (#3) #4}
\def\NCA{\em Nuovo Cimento}
\def\NIM{\em Nucl. Instrum. Methods}
\def\NIMA{{\em Nucl. Instrum. Methods} {\bf A}}
\def\NPB{{\em Nucl. Phys.}   {\bf B}}
\def\PLB{{\em Phys. Lett.}   {\bf B}}
\def\PRL{\em Phys. Rev. Lett.}
\def\PRD{{\em Phys. Rev.}    {\bf D}}
\def\ZPC{{\em Z. Phys.}      {\bf C}}
\def\EJC{{\em Eur. Phys. J.} {\bf C}}
\def\CPC{\em Comp. Phys. Commun.}

\begin{titlepage}
 
\noindent
DESY 02-224  \hfill  ISSN 0418-9833 \\
January 2003

\vspace*{3cm}
 
\begin{center}
  \Large {\boldmath \bf Isolated electrons and muons in
    events with missing transverse momentum at HERA}

  \vspace*{1cm}
    {\Large H1 Collaboration} 
\end{center}

\begin{abstract}
\noindent
A search for events with a high energy isolated electron or muon and
missing transverse momentum has been performed at the electron--proton
collider HERA using an integrated luminosity of $13.6$ pb$^{-1}$ in
$e^-p$ scattering and $104.7$ pb$^{-1}$ in $e^+p$ scattering. Within
the Standard Model such events are expected to be mainly due to $W$
boson production with subsequent leptonic decay. In $e^-p$
interactions one event is observed in the electron channel and none in
the muon channel, consistent with the expectation of the Standard
Model. In the $e^+p$ data a total of $18$ events are seen in the
electron and muon channels compared to an expectation of $12.4 \pm
1.7$ dominated by $W$ production ($9.4 \pm 1.6$). Whilst the overall
observed number of events is broadly in agreement with the number
predicted by the Standard Model, there is an excess of events with
transverse momentum of the hadronic system greater than $25$ GeV with
$10$ events found compared to $2.9 \pm 0.5$ expected. The results are
used to determine the cross section for events with an isolated
electron or muon and missing transverse momentum.

\end{abstract}

\vfill
\begin{center}
To be submitted to \PLB
\end{center}

\end{titlepage}

\begin{flushleft}

V.~Andreev$^{24}$,             
B.~Andrieu$^{27}$,             
T.~Anthonis$^{4}$,             
A.~Astvatsatourov$^{35}$,      
A.~Babaev$^{23}$,              
J.~B\"ahr$^{35}$,              
P.~Baranov$^{24}$,             
E.~Barrelet$^{28}$,            
W.~Bartel$^{10}$,              
S.~Baumgartner$^{36}$
% ...

\bigskip{\it
 $ ^{1}$ I. Physikalisches Institut der RWTH, Aachen, Germany$^{ a}$ \\
% ...
$ ^{4}$ Inter-University Institute for High Energies ULB-VUB, Brussels;
          Universiteit Antwerpen (UIA), Antwerpen; Belgium$^{ c}$ \\
% ...
$ ^{10}$ DESY, Hamburg, Germany \\
% ...
$ ^{23}$ Institute for Theoretical and Experimental Physics,
          Moscow, Russia$^{ l}$ \\
 $ ^{24}$ Lebedev Physical Institute, Moscow, Russia$^{ e}$ \\
% ...
$ ^{27}$ LPNHE, Ecole Polytechnique, IN2P3-CNRS, Palaiseau, France \\
 $ ^{28}$ LPNHE, Universit\'{e}s Paris VI and VII, IN2P3-CNRS,
          Paris, France \\
% ...
$ ^{35}$ DESY, Zeuthen, Germany \\
 $ ^{36}$ Institut f\"ur Teilchenphysik, ETH, Z\"urich, Switzerland$^{ j}$ \\
% ...
$ ^a$ Supported by the Bundesministerium f\"ur Bildung und Forschung, FRG,
      under contract numbers 05 H1 1GUA /1, 05 H1 1PAA /1, 05 H1 1PAB /9,
      05 H1 1PEA /6, 05 H1 1VHA /7 and 05 H1 1VHB /5 \\
% ...
$ ^c$ Supported by FNRS-FWO-Vlaanderen, IISN-IIKW and IWT \\
% ...
$ ^e$ Supported by the Deutsche Forschungsgemeinschaft \\
% ...
$ ^j$ Supported by the Swiss National Science Foundation \\
% ...
$ ^l$ Partially Supported by Russian Foundation
      for Basic Research, grant    no. 00-15-96584 \\
}

\end{flushleft}
\newpage

\pagestyle{plain}

\section{Introduction}
\label{sec:intro}

The HERA collaborations H1 and ZEUS have previously reported
\cite{origisol,Adloff:1998aw,Breitweg:1999ie} the observation of
events with an isolated high energy lepton and missing transverse
momentum in $e^{+}p$ collisions recorded during the period 1994--1997.
The dominant Standard Model (SM) contribution to this topology is real
$W$ boson production with subsequent leptonic decay. Such events can
also be a signature of new phenomena beyond the Standard Model
\cite{bsm}.  H1 has reported \cite{Adloff:1998aw} one $e^-$ event and
$5$ $\mu^{\pm}$ events compared to expectations from the Standard
Model of $2.4 \pm 0.5$ and $0.8 \pm 0.2$ for the $e^{\pm}$ and
$\mu^{\pm}$ channels respectively, with $W$ contributions of $1.65 \pm
0.47$ ($e$) and $0.53 \pm 0.11$ ($\mu$). For the same data taking
period ZEUS has reported \cite{Breitweg:1999ie} $3$ ($0$) $e^{\pm}$
($\mu^{\pm}$) events compared to an expectation of $2.1$ ($0.8$) $W$
events and $1.1 \pm 0.3$ ($0.7 \pm 0.2$) events from other processes.
In the present paper a search for events with isolated
electrons\footnote{In this paper ``electron'' refers generically to
  both electrons and positrons.  Where distinction is required the
  terms $e^-$ and $e^+$ are used.} or muons and missing transverse
momentum is performed in an extended phase space and with improved
background rejection. The complete HERA I data sample (1994--2000) is
analysed here. This corresponds to an integrated luminosity of $118.4$
pb$^{-1}$, which represents a factor of three increase with respect to
the previous published result.

This paper is organised as follows. Section \ref{sec:monte} describes
the SM processes that contribute to the signal and to the background.
Section \ref{sec:exper} describes the H1 detector and experimental
conditions. Section \ref{sec:lephad} outlines the lepton
identification criteria and the reconstruction methods for the
hadronic final state. The selection requirements for the electron and
muon channels are described in section~\ref{sec:lepsel}.  Studies of
background processes are presented in section~\ref{sec:lepback}.
Section \ref{sec:expsyst} deals with systematic uncertainties and
section \ref{sec:results} presents the results of the analysis
including the numbers of events seen, the kinematics of the selected
events and the measured cross sections.  The results of a search for
$W$ production in the hadronic decay channel are given in
section~\ref{sec:hadchan}. The paper is briefly summarised in
section~\ref{sec:summary}.

\section{Standard Model Processes}
\label{sec:monte}

The processes within the Standard Model that are expected to lead to a
final state containing an isolated electron or muon and missing
transverse momentum, due to penetrating particles escaping detection
in the apparatus, are described in detail in \cite{Adloff:1998aw} and
are only briefly outlined in this section.  The processes are called
``signal'' if they produce events which contain a genuine isolated
electron or muon and genuine missing transverse momentum in the final
state. The processes are defined as ``background'' if they contribute
to the selected sample through misidentification or mismeasurement.
For the background processes, a fake lepton, fake missing transverse
momentum or both can be reconstructed and may lead to the topology of
interest. The following processes are considered.

\begin{itemize}
  
\item[] {\boldmath \bf $W$ production}: $ep\rightarrow e W^\pm X$ or
  $ep\rightarrow \nu W^\pm X$ (signal)
  
  Real $W$ production in electron proton collisions with subsequent
  leptonic decay $W\rightarrow l\nu$, proceeding via photoproduction,
  is the dominant SM process that produces events with prominent high
  $P_T$ isolated leptons and missing transverse momentum. $W$ bosons
  are predicted to be produced mainly in resolved photon interactions,
  in which the $W$ typically has small transverse momentum, whilst in
  direct photon interactions the $W$ transverse momentum may be
  larger.
  
  In this paper, the SM prediction for $W$ production via
  $ep\rightarrow e W^\pm X$ is calculated by using a next to leading
  order (NLO) Quantum Chromodynamics (QCD) calculation
  \cite{Diener:2002if} in the framework of the EPVEC
  \cite{Baur:1991pp} event generator. Each event generated by EPVEC
  according to its default LO cross section is weighted by a factor
  dependent on the transverse momentum and rapidity of the $W$
  \cite{spiraprivate}, such that the resulting cross section
  corresponds to the NLO calculation. The ACFGP \cite{Aurenche:1992sb}
  parameterisation is used for the photon structure and the CTEQ4M
  \cite{Lai:1996mg} parton distribution functions are used for the
  proton. The renormalisation scale is taken to be equal to the
  factorisation scale and is fixed to the $W$ mass. Final state parton
  showers are simulated using the PYTHIA framework \cite{epvecmcw}.
    
  The NLO corrections are found to be of the order of $30\%$ at low
  $W$ transverse momentum (resolved photon interactions) and typically
  $10\%$ at high $W$ transverse momentum (direct photon interactions)
  \cite{Diener:2002if}.  The NLO calculation reduces the theory error
  to $15\%$ (from $30\%$ at leading order).
  
  The charged current process $ep\rightarrow \nu W^\pm X$ is
  calculated with EPVEC \cite{Baur:1991pp} and found to contribute
  less than $7\%$ of the predicted signal cross section.
  
  The total predicted $W$ production cross section amounts to $1.1$ pb
  for an electron--proton centre of mass energy of $\sqrt{s}=300$ GeV
  and $1.3$ pb for $\sqrt{s}=318$ GeV.
  
\item[] {\boldmath \bf $Z$ production} : $ep\to eZ(\rightarrow \nu
  \bar{\nu})X$ (signal)
  
  A small number of signal events may be produced by $Z$ production
  with subsequent decay to neutrinos. The outgoing electron from this
  reaction can scatter into the detector yielding the isolated lepton
  in the event while genuine missing transverse momentum is produced
  by the neutrinos. This process is calculated with the EPVEC
  generator and found to contribute less than $3\%$ of the predicted
  signal cross section.
  
\item[] {\boldmath \bf Charged Current (CC) processes} : $ep\to\nu X$
  (background)
  
  A CC deep inelastic event can mimic the selected topology if a
  particle in the hadronic final state or a radiated photon is
  interpreted as an isolated lepton.  The generator DJANGO
  \cite{django} is used to calculate this contribution to the
  background.
  
\item[] {\boldmath \bf Neutral Current (NC) processes} : $ep\to eX$
  (background)
  
  The scattered electron in a NC deep inelastic event yields an
  isolated high energy lepton, but measured missing transverse
  momentum can only be produced by fluctuations in the detector
  response or by undetected particles due to limited geometrical
  acceptance.  The generator RAPGAP \cite{Jung:1993gf} is used to
  calculate this contribution to the background.
  
\item[] {\boldmath \bf Photoproduction of jets} : $\gamma p\to X$
  (background)
  
  The generator PYTHIA \cite{Sjostrand:1993yb} is used to calculate
  the contribution from hard scattering photoproduction processes.
  Background from this process may occur if a particle from the
  hadronic final state is interpreted as an isolated lepton and
  missing transverse momentum is measured due to fluctuations in the
  detector response or limited geometrical acceptance.
  
\item[] {\boldmath \bf Lepton pair (LP) production} : $ep \rightarrow
  e~l^+ l^-X$ (background)
 
  Lepton pair production can mimic the selected topology if one lepton
  escapes detection and measurement errors cause apparent missing
  momentum. The generator GRAPE 1.1 \cite{grape}, based on a full
  calculation of electroweak diagrams, is used. The dominant
  contribution is due to photon--photon processes and is cross-checked
  with the LPAIR \cite{lpair} generator. Internal photon conversions
  are also calculated. $Z$ production and its subsequent decay into
  charged leptons is also included in GRAPE.  This contribution is
  found to be negligible.

\end{itemize}

In order to determine signal acceptances and background contributions,
the detector response to events produced by the above programs is
simulated in detail using a program based on GEANT \cite{Brun:1987ma}.
The simulated events are then subjected to the same reconstruction and
analysis chain as the data.

\section{Experimental Conditions} 
\label{sec:exper}

Results are presented for the $37.0$ pb$^{-1}$ of $e^+p$ data taken in
1994--1997 at an electron--proton centre of mass energy of
$\sqrt{s}=300$ GeV , the $13.6$ pb$^{-1}$ of $e^-p$ data (1998--1999,
$\sqrt{s}=318$ GeV) and the $67.7$ pb$^{-1}$ of $e^+p$ data
(1999--2000, $\sqrt{s}=318$ GeV).

A detailed description of the H1 detector can be found in
\cite{Abt:hi}. Only those components of particular
importance to this analysis are described here.

The inner tracking system consisting of central and
forward\footnote{The forward direction and the positive $z$--axis are
  taken to be that of the proton beam direction. All polar angles are
  defined with respect to the positive $z$--axis.} tracking detectors
(drift chambers) is used to measure charged particle trajectories and
to determine the interaction vertex.  A solenoidal magnetic field
allows the measurement of the particle transverse momenta.

Electromagnetic and hadronic final state particles are absorbed in a
highly segmented Liquid Argon (LAr) calorimeter \cite{Andrieu:1993kh}.
The calorimeter is $5$ to $8$ interaction lengths deep depending on
the polar angle of the particle. A lead--fibre calorimeter (SpaCal) is
used to detect backward going electrons and hadrons.

The LAr calorimeter is surrounded by a superconducting coil with an
iron return yoke instrumented with streamer tubes. Tracks of muons,
which penetrate beyond the calorimeter, are reconstructed from their
hit pattern in the streamer tubes. The instrumented iron is also used
as a backing calorimeter to measure the energy of hadrons that are not
fully absorbed in the LAr calorimeter.

In the forward region of the detector a set of drift chamber layers
(the forward muon system) detects muons and, together with an iron
toroidal magnet, allows a momentum measurement. Around the beam pipe,
the plug calorimeter measures hadronic activity at low polar angles.

The LAr calorimeter provides the main trigger for events with high
transverse momentum. The trigger efficiency is $\simeq 98\%$ for
events with an electron which has transverse momentum above $10$ GeV.
For events with high missing transverse momentum, determined from an
imbalance in transverse momentum measured in the calorimeter $P^{\rm
  calo}_{T}$, the trigger efficiency is $\simeq 98\%$ when $P^{\rm
  calo}_{T}>25$ GeV and is $\sim 50\%$ when $P^{\rm calo}_{T} =12$ GeV
\cite{Adloff:1999ah}. Events may also be triggered by a pattern
consistent with a minimum ionising particle in the muon system in
coincidence with tracks in the tracking detectors.

\section{Lepton Identification and Hadronic Reconstruction}
\label{sec:lephad}

An electron candidate is defined \cite{bruelthesis} by the presence of
a compact and isolated electromagnetic cluster of energy in the LAr
calorimeter, with the requirement of an associated track having an
extrapolated distance of closest approach to the cluster of less than
$12$ cm.  Electrons found in regions between calorimeter modules
containing large amounts of inactive material are excluded
\cite{Adloff:1999ah}.  The energy of the electron candidate is
measured from the calorimeter cluster.  The additional energy allowed
within a cone of radius $1$ in pseudorapidity--azimuth
($\eta$--$\phi$) space around the electron candidate is required to be
less than $3\%$ of the energy attributed to the electron candidate.
The efficiency of electron identification is established using NC
events and is greater than $98\%$ \cite{Adloff:1999ah}.

A muon candidate is identified by a track in the forward muon system
or a charged track in the inner tracking system associated with a
track segment or an energy deposit in the instrumented iron. The muon
momentum is measured from the track curvature in the solenoidal or
toroidal magnetic field. A muon candidate may have no more than $8$
GeV deposited in the LAr calorimeter in a cone of radius $0.5$ in
($\eta$--$\phi$) space associated with its track. The efficiency to
identify muons is established using elastic LP events
\cite{boristhesis} and is greater than $90\%$.

Identified leptons are characterised by the following variables, where
$l$ represents $e$ or $\mu$:

\begin{itemize}

\item $P^l_T$, the transverse momentum of an identified muon or electron;

\item $\theta_l$, the polar angle of the muon or electron.

\end{itemize}

\begin{figure}[ht]
  \setlength{\unitlength}{1cm}
  \begin{picture}(12.0,8.0) 
    \put(-1.0,0.0) {\epsfig{file=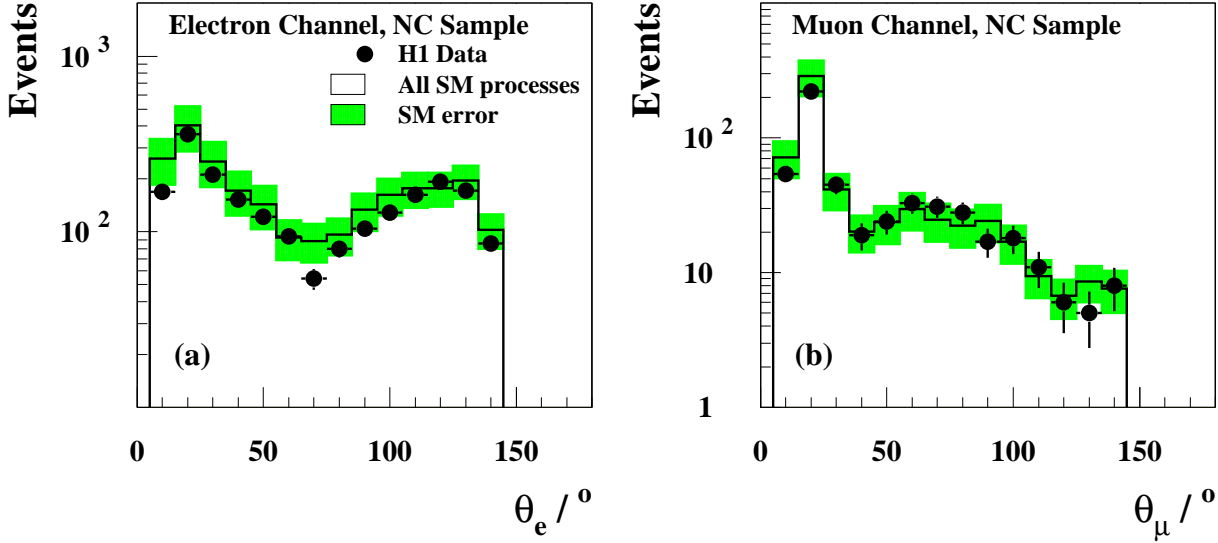,
        width=1.1\textwidth,bbllx=1pt,bblly=234pt,bburx=560pt,bbury=560pt}}
    \put(1.25,3.5) {{\bf(a)}}
    \put(9.5,3.5) {{\bf(b)}}
  \end{picture}
  \caption{Distributions of lepton polar angle of a second reconstructed
    electron (a) and a reconstructed muon (b) for NC events. The
    combined SM expectation is shown as an open histogram. The total
    error on the SM expectation (see section \ref{sec:expsyst}) is
    given by the shaded band.  }
  \label{fig:lepidnc} 
\end{figure}

In order to check that the probability to misidentify a particle as an
electron or muon is well described by the simulation, a sample of NC
events is used, in which a second electron or a muon is found in the
event. In the majority of cases this second lepton results from the
misidentification of a hadron from the final state.  The second lepton
in the event must pass the same criteria as described above, except
for the upper limit on the calorimeter energy within a cone associated
with its track. The study is performed requiring the reconstructed
electrons or muons to have $P_T^l>10$ GeV.  From a total NC sample of
121408 events, 2087 events with a second identified electron and 520
events with a reconstructed muon are selected by this procedure.
Figure \ref{fig:lepidnc}a shows the polar angle distribution of the
electron with the second highest transverse momentum and figure
\ref{fig:lepidnc}b shows the polar angle distribution of reconstructed
muons. The distributions are described by the simulation within the
uncertainties, demonstrating that the misidentification of a particle
as an electron or muon is well understood.

The hadronic final state (HFS) is measured by combining calorimeter
energy deposits with low momentum tracks as described in
\cite{Adloff:1999ah}.  Identified isolated electrons or muons are
excluded from the HFS. The calibration of the hadronic energy scale is
made by comparing the transverse momentum of the precisely measured
scattered electron to that of the HFS in a large NC event sample. The
transverse momentum of the hadronic system is:

\begin{itemize}

\item $P^X_T$, which includes all reconstructed particles apart from
  identified isolated leptons.

\end{itemize}

The isolation of identified leptons with respect to jets or other
tracks in the event is quantified using:

\begin{itemize}
  
\item their distance $D_{jet}$ from the axis of the closest hadronic
  jet in $\eta$--$\phi$ space.  For this purpose jets, excluding
  identified leptons, are reconstructed using an inclusive $k_T$
  algorithm \cite{Ellis:tq,Catani:1993hr,Adloff:1998ni} and are
  required to have transverse momentum greater than $5$ GeV. If there
  is no such jet in the event, $D_{jet}$ is defined with respect to
  the polar and azimuthal angles of the hadronic final state;
  
\item their distance $D_{track}$ from the closest track in
  $\eta$--$\phi$ space, where all tracks with a polar angle greater
  than $10^{\circ}$ and transverse momentum greater than $0.15$ GeV
  are considered.

\end{itemize}

The following quantities are sensitive to the presence of high energy
undetected particles and/or can be used to reduce the main background
contributions.

\begin{itemize}
  
\item $P^{\rm calo}_{T}$, the net transverse momentum measured from
  all energy deposits recorded in the calorimeter.
  
\item $P^{\rm miss}_{T}$, the total missing transverse momentum
  reconstructed from all observed particles (electrons, muons and
  hadrons). $P^{\rm miss}_{T}$ differs most from $P^{\rm calo}_{T}$ in
  the case of events with muons, since they deposit little energy in
  the calorimeter.
  
\item $\frac{V_{\rm ap}}{V_{\rm p}}$, a measure of the azimuthal
  balance of the event.  It is defined as the ratio of the
  anti--parallel to parallel components of the measured calorimetric
  transverse momentum, with respect to the direction of the
  calorimetric transverse momentum \cite{Adloff:1999ah}.  Events with
  one or more high $p_T$ particles that do not deposit much energy in
  the calorimeter ($\mu$, $\nu$) generally have low values of
  $\frac{V_{\rm ap}}{V_{\rm p}}$.
  
\item $\delta_{\rm miss}=2E_{e}-\sum_{i} E_i(1-\cos \theta_i)$, where
  $E_i$ and $\theta_i$ denote the energy and polar angle of each
  particle in the event detected in the main detector
  ($\theta_e<176^\circ$) and $E_e$ is the electron beam energy. For an
  event where only momentum in the proton direction is undetected
  $\delta_{\rm miss}$ is zero.

\item $\Delta \phi_{l-X}$, the difference in azimuthal angle between
  the lepton and the direction of  $P^X_T$. NC events typically
  have values of $\Delta \phi_{l-X}$ close to $180^\circ$.
  
\item ${\zeta}^{2}_{e}=4 E^{'}_{e}E_e \cos^2 \theta_e/2$, where
  $E^{'}_{e}$ is the energy of the final state electron. For NC
  events, where the scattered electron is identified as the isolated
  high transverse momentum electron, ${\zeta}^{2}_{e}$ is equal to the
  four momentum transfer squared $Q^2$. Since the NC cross section
  falls steeply with $Q^2$, these events generally have small values
  of ${\zeta}^{2}_{e}$. Conversely, electrons from $W$ decay generally
  have high values of ${\zeta}^{2}_{e}$.

\end{itemize}

\section{Selection Criteria}
\label{sec:lepsel}

The published H1 observation \cite{Adloff:1998aw} using 1994--1997
$e^+p$ data was based on the selection of a sample of events with
$P_T^{\rm calo}>25$ GeV. This experimental cut mainly selected charged
current events in a phase space where the trigger efficiency is high.
In the selected events all isolated charged tracks with transverse
momentum above $10$ GeV were identified as electrons or muons.

In the present paper the $P_T^{\rm calo}$ cut has been lowered to $12$
GeV, taking advantage of the improved understanding of trigger
efficiencies with increased luminosity and more sophisticated
background rejection. The analysis extends the phase space towards
lower missing transverse momentum for the electron channel ($P_T^{\rm
  calo} \simeq P_T^{\rm miss}$) and towards lower $P_T^X$ for the muon
channel ($P_T^{\rm calo} \simeq P_T^{X}$). The lepton identification
has also been improved and extended in the forward direction. The
increased phase space and increased luminosity allow the comparison
with the SM predictions to be made differentially and with improved
precision. Further details of the analysis can be found in
\cite{nickthesis,mireillethesis}.

The selection criteria for both channels are summarised in
table~\ref{cutsb}. The dominant background in the electron channel is
due to NC and CC events. To reduce the NC background, events with NC
topology (azimuthally balanced, with the lepton and the hadronic
system 
back-to-back
in the transverse plane) are rejected.
For low values of $P_{T}^{\rm calo}$, where the NC background is
largest, a requirement on $\zeta^{2}_e$ is imposed. A requirement that
the lepton candidate be isolated from the hadronic final state is
imposed to reject CC events.  Events which have, in addition to an
isolated electron, one or more isolated muons are not considered in
the electron channel, but may contribute in the muon channel. The
dominant backgrounds in the muon channel are inelastic muon pair
production and CC or photoproduction events which contain a
reconstructed muon. The final muon sample is selected by rejecting
azimuthally balanced events and events where more than one muon is
observed.

Following the selection criteria described above, the overall
efficiency to select SM $W \to e \nu$ events is $41\%$ and to select
SM $W \to \mu \nu$ events is $14\%$. The main difference in efficiency
between the two channels is due to the cut on $P_{T}^{\rm calo}$,
which for muon events acts as a cut on $P_{T}^{X}$ because the muon
deposits little energy in the calorimeter. There is thus almost no
efficiency in the muon channel for $P_{T}^{X}<12$ GeV.  For values of
$P_{T}^{X}>25$ GeV the efficiencies of the two channels are compatible
at $\sim 40\%$.

\begin{table}[h]
\begin{center}
\begin{tabular}{|c||c|c|} \hline
 Variable  & Electron  & Muon \\ \hline
\hline
$\theta_l$ &  \multicolumn{2}{c|}{$5^\circ<\theta_l<140^\circ$}\\
\hline
$P_T^l$ & \multicolumn{2}{c|}{$ >10$ GeV}\\
\hline
$P_T^{\rm calo}$ & \multicolumn{2}{c|}{$> 12$ GeV}\\
\hline                                                   
$P_T^{\rm miss}$ & \multicolumn{2}{c|}{$> 12$ GeV}\\
\hline
$P_T^X$ &  --  &$> 12$ GeV\\ 
\hline
$D_{jet}$ & \multicolumn{2}{c|}{$>$ $1.0$}\\
\hline
$D_{track}$ & $> 0.5$ for $\theta_e \ge 45^\circ$ &  $> 0.5$ \\
\hline
$\zeta^2_l$ & $>5000$ GeV$^2$ for $P_T^{\rm calo} < 25$ GeV & --\\
\hline
$\frac{V_{ap}}{V_p}$ & $< 0.5$ ($< 0.15$ for $P_T^e < 25$ GeV) & $< 0.5$ ($<$ $0.15$ for $P_T^{\rm calo} < 25$ GeV)\\
\hline
$\Delta\phi_{l-X}$ & $< 160^\circ$ & $< 170^\circ$ \\
\hline
\# isolated $\mu$ & $0$ & $1$\\
\hline
$\delta_{\rm miss}$ & $ > 5$ GeV $^\dagger$ & -- \\
\hline
\end{tabular}
\begin{flushright}
  $^\dagger$ {\it if only one $e$ candidate is detected, which has the same charge as the beam lepton.}
\end{flushright}
\end{center}
\caption{Selection requirements for the electron and muon channels.}
\label{cutsb}
\end{table}

\section{Background Studies}
\label{sec:lepback}

To verify that the backgrounds (see section \ref{sec:monte}) that
contribute to the two channels are well understood, alternative event
samples, each enriched in one of the important background processes,
are compared with simulations. For both channels these event samples
have the same basic phase space definition ($\theta_l$, $P_T^l$,
$P_T^{\rm calo}$) as the main analysis. It should be noted that these
selections do not explicitly reject signal events, which may be
present in the enriched samples.

The two background enriched samples in the electron channel, defined
within the phase space $5^\circ<\theta_e<140^\circ$, $P_T^e$ $> 10$
GeV and $P_T^{\rm calo}> 12$ GeV, are selected with the following
additional requirements.

\begin{itemize}

\item[]{\bf NC enriched sample}\\ A NC dominated electron sample is
  selected by requiring $D_{jet}> 1.0$. The events in this channel
  mainly contain genuine electron candidates, but with missing
  transverse momentum arising from mismeasurement.
  
\item[]{\bf CC enriched sample}\\ A CC dominated sample is obtained by
  rejecting events with an isolated muon and applying cuts to suppress
  the NC component. These criteria are $\zeta^2_e\ge 2500$ GeV$^2$,
  $\frac{V_{ap}}{V_p}\le 0.15$, $\delta_{\rm miss}> 5$ GeV and
  $\Delta\phi_{e-X}<$ 160$^\circ$. In this sample the missing
  transverse momentum is genuine, but an electron candidate is usually
  falsely identified.

\end{itemize}

The two samples designed to study the backgrounds in the muon channel,
defined within the same phase space $5^\circ<\theta_{\mu}<140^\circ$,
$P_T^{\mu}$ $> 10$~GeV and $P_T^{\rm calo}> 12$~GeV, are selected with
the following additional requirements.

\begin{itemize}
  
\item[]{\bf LP enriched sample}\\ A sample of events predominantly
  from the two photon process is selected by requiring at least one
  isolated muon and $\frac{V_{ap}}{V_p}\le 0.2$ to suppress
  photoproduction events.
    
\item[]{\bf CC enriched sample} A sample dominated by CC events is
  selected by requiring $\frac{V_{ap}}{V_p}\le 0.15$ and requiring at
  least one muon candidate that need not be isolated.  This selection
  tests fake or real muons observed in events with genuine missing
  $P_T$.
  
\end{itemize}

\begin{figure}[ht]
  \setlength{\unitlength}{1cm}
  \begin{picture}(12.0,15.0)
    \put(-1.0,0.0) {\epsfig{file=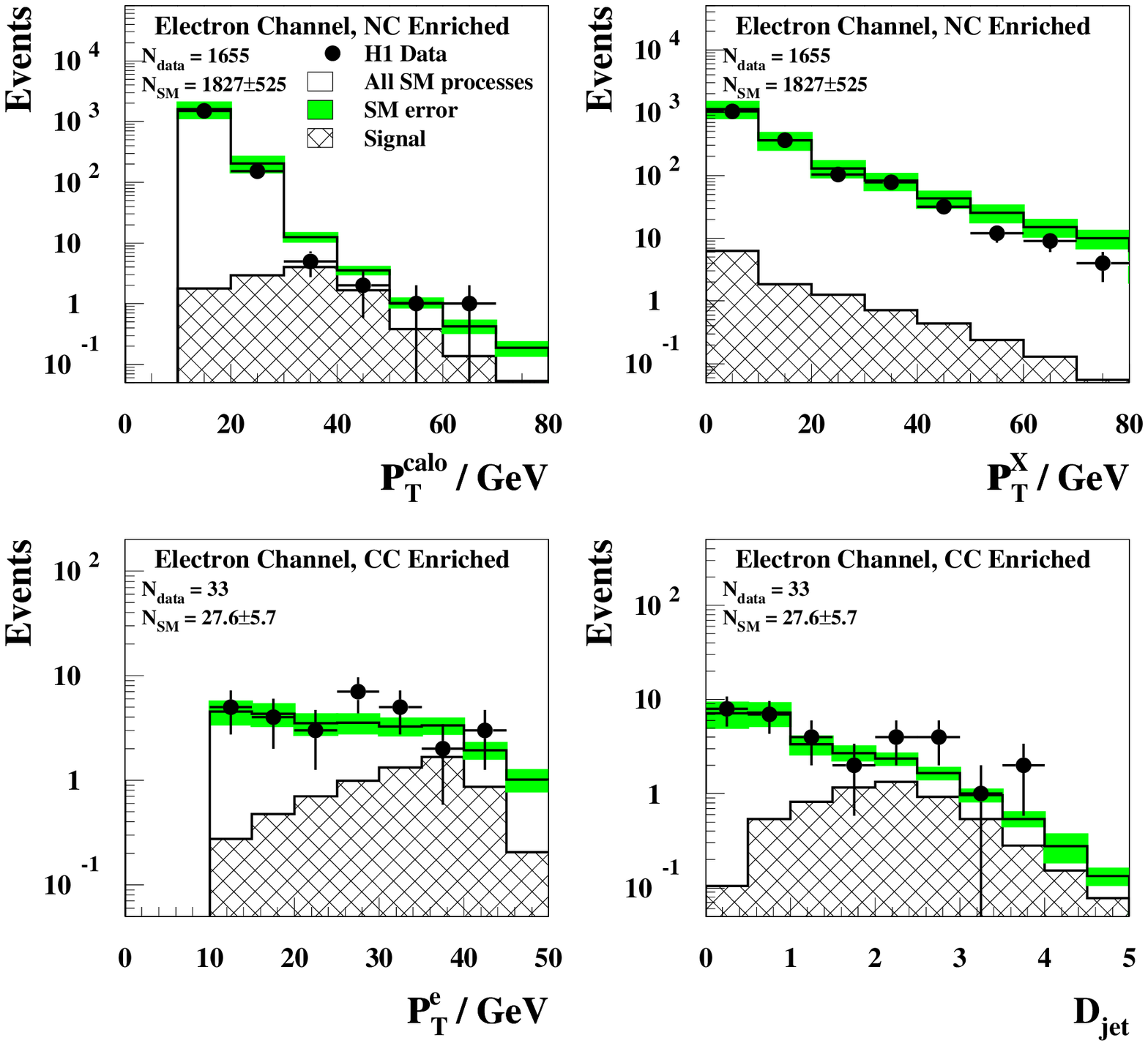,width=1.1\textwidth}}
    \put(2.15,3.25) {{\bf(c)}}
    \put(9.65,3.35) {{\bf(d)}}
    \put(1.85,10.7) {{\bf(a)}}
    \put(9.65,10.7) {{\bf(b)}}
  \end{picture}
  \caption{The $e^+p$ data selected in the NC enriched sample
    (a,b) and in the CC enriched sample (c,d) in the electron channel
    compared with the combined SM expectation (open histogram). The
    total error on the SM expectation is given by the shaded band. The
    ``signal'' component of the SM expectation is given by the hatched
    histogram. $\rm N_{data}$ is the total number of data events
    observed for each sample. $\rm N_{SM}$ is the total SM
    expectation.}
  \label{fig:econnc} 
\end{figure}

\begin{figure}[ht]
  \setlength{\unitlength}{1cm}
  \begin{picture}(12.0,15.0)
    \put(-1.0,0.0) {\epsfig{file=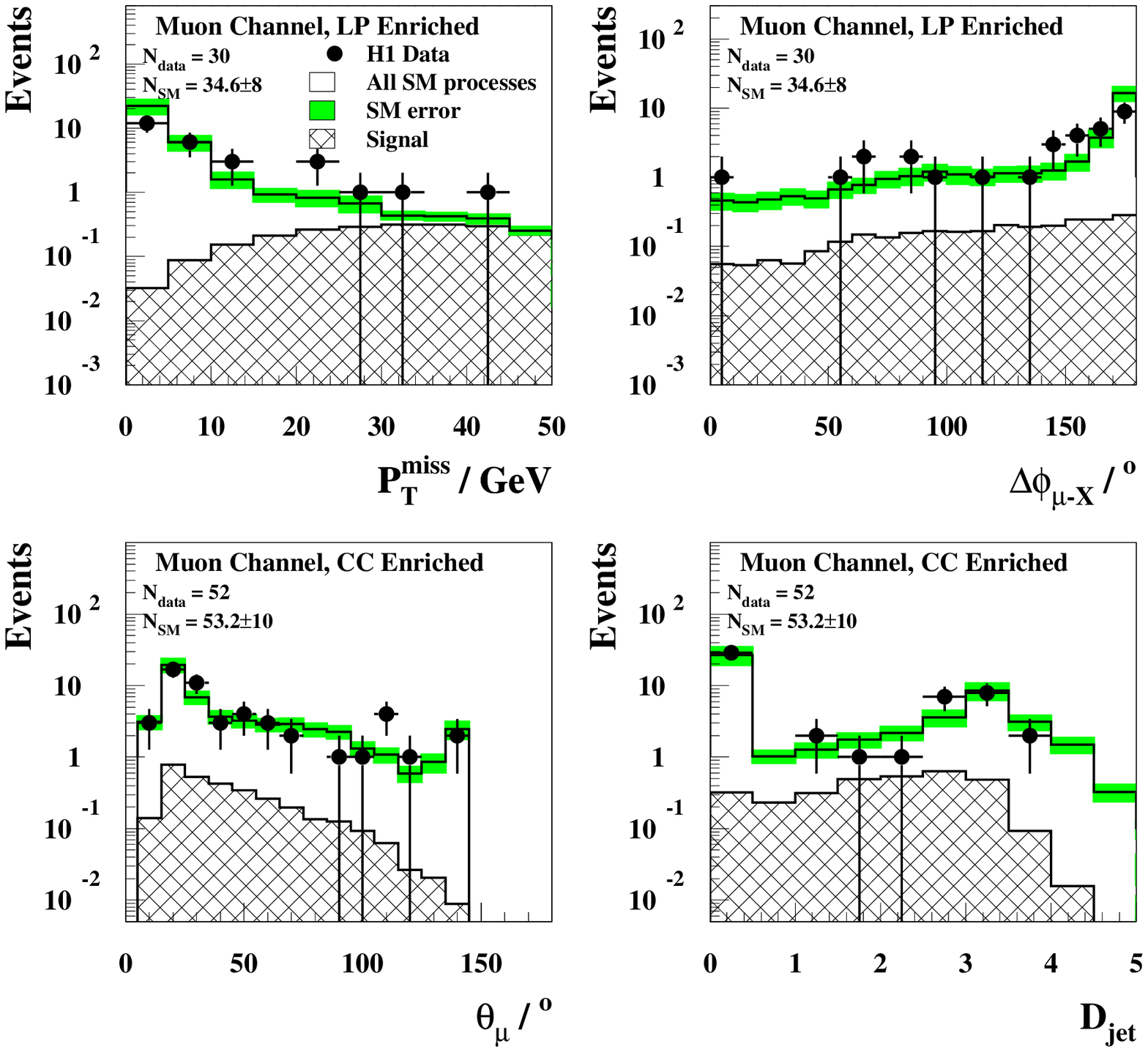,width=1.1\textwidth}}
    \put(1.45,3.35) {{\bf(c)}}
    \put(9.65,3.35) {{\bf(d)}}
    \put(1.45,10.7) {{\bf(a)}}
    \put(9.65,10.7) {{\bf(b)}}
  \end{picture}
  \caption{The $e^+p$ data selected in the LP enriched sample
    (a,b) and in the CC enriched sample (c,d) in the muon channel
    compared with the combined SM expectation (open histogram). The
    total error on the SM expectation is given by the shaded band. The
    ``signal'' component of the SM expectation is given by the hatched
    histogram. $\rm N_{data}$ is the total number of data events
    observed for each sample. $\rm N_{SM}$ is the total SM
    expectation.}
  \label{fig:mconlp} 
\end{figure}

The distributions of all quantities used in these selections are well
described in both shape and normalisation by the SM expectation in
regions where there is little contribution from $W$ production. This
gives us confidence that the backgrounds are described within the
uncertainty.  Example distributions of the background enriched event
samples for the $e^+p$ data are shown in figure~\ref{fig:econnc} for
the electron channel and in figure~\ref{fig:mconlp} for the muon
channel.  Also included in the figures are the SM expectations from
all processes together and the signal expectation alone. Agreement is
also obtained between the data and the simulation in all distributions
for the $e^-p$ data sample.

\section{Systematic Uncertainties}
\label{sec:expsyst}

The systematic uncertainties on quantities which influence the SM
expectation and the measured cross section (see section
\ref{sec:crosssec}) are presented in this section and discussed in
more detail in \cite{Adloff:1999ah,mireillethesis}.  The uncertainties
on the signal expectation and the acceptance used in the cross section
calculation are determined by varying experimental quantities by $\pm$
$1$ standard deviation and recalculating the cross section or
expectation. The experimental uncertainties are listed below and the
corresponding variation of the cross section is given in
table~\ref{systematics}.

\begin{itemize}

\item {\bf Leptonic quantities}\\ The uncertainties on the $\theta_l$
  and the $\phi_l$ measurements are $3$ mrad and $1$ mrad
  respectively. The electron energy scale uncertainty is $3\%$. The
  muon energy scale uncertainty is $5\%$.
  
\item {\bf Hadronic quantities}\\ The uncertainties on the $\theta$
  and $\phi$ measurements of the hadronic final state are both $20$
  mrad.  The hadronic energy scale uncertainty is $4\%$. The error on
  the measurement of $\frac{V_{\rm ap}}{V_{\rm p}}$ is $\pm 0.02$.
  
\item {\bf Triggering / Identification}\\ The electron finding
  efficiency has an uncertainty of $2\%$. The muon finding efficiency
  has an error of $5\%$ in the central ($\theta_{\mu}>12.5^\circ$)
  region and $15\%$ in the forward ($\theta_{\mu}<12.5^\circ$) region.
  The uncertainty on the track reconstruction efficiency is $3\%$.
  The uncertainty on the trigger efficiency for the muon channel
  varies from $16\%$ at $P^X_T=12$~GeV to $2\%$ at $P^X_T>40$~GeV.

\item {\bf Luminosity}\\ The luminosity measurement has an uncertainty
 of $1.5\%$.
 
\item {\bf Model}\\ A $10\%$ uncertainty on the model dependence of
  the acceptance is estimated by comparing the results obtained with
  two further generators which produce $W$ bosons with different
  kinematic distributions from those of EPVEC. The generators used are
  an implementation of $W$ production within PYTHIA and ANOTOP, an
  ``anomalous top production'' generator, using the matrix elements of
  the complete process $e + q \rightarrow e + t \rightarrow e + b + W$
  as obtained from the CompHEP program \cite{Boos:1994xb}.

\end{itemize}

Contributions from background processes, modelled using RAPGAP, DJANGO
and GRAPE, are attributed $30\%$ systematic errors determined from the
level of agreement observed between the simulations and the control
samples (see section \ref{sec:lepback}).  The uncertainties associated
with lepton misidentification and the production of fake missing
transverse momentum are included in these errors.

A theoretical uncertainty of $15\%$ is quoted for the predicted
contributions from signal processes (predominantly SM $W$ production).
This is due mainly to uncertainties in the parton distribution
functions and the scales at which the calculation is performed
\cite{Diener:2002if}.

\begin{table}[h]
  \begin{center}
    \begin{tabular}{|c||c|c|} \hline
      Source of systematic uncertainty  &  \multicolumn{2}{c|}{Error on measured cross section}\\
      \cline{2-3}      & $P_T^X < 25$ GeV & $P_T^X > 25$ GeV\\
      \hline \hline
      Leptonic quantities       & $\pm 0.6\%$  & $\pm 0.6\%$ \\ \hline
      Hadronic quantities       & $\pm 3.3\%$  & $\pm 8.3\%$ \\ \hline
      Triggering/Identification & $\pm 3.7\%$  & $\pm 4.7\%$ \\ \hline
      Luminosity                & $\pm 1.5\%$  & $\pm 1.5\%$ \\ \hline
      Model Uncertainty         & $\pm 10\%$   & $\pm 10\%$ \\ \hline
    \end{tabular}
  \end{center}
  \caption{Summary of experimental systematic errors on the measured cross section in two regions of $P_T^X$.}
  \label{systematics}
\end{table}

\section{Results}
\label{sec:results}

For the $e^-p$ data sample one event is observed in the electron
channel. The kinematics of the event are listed in table~\ref{evkin}.
No events are observed in the muon channel. This compares well to the
SM expectations of $1.69 \pm 0.22$ events in the electron channel and
$0.37 \pm 0.06$ in the muon channel.

In the $e^+p$ data sample $10$ candidate events are observed in the
electron channel compared to $7.2 \pm 1.2$ expected from signal
processes and $2.68 \pm 0.49$ from background sources.  One candidate
event in the electron channel is observed to contain an $e^-$. This
event was first reported and discussed in \cite{Adloff:1998aw}.  Four
of the other candidate events contain an $e^+$.  The charges of the
electrons in the remaining five events are unmeasured since the
electrons are produced at low polar angles and they shower in material
in the tracking detectors.  In the muon channel $8$ candidate events
are observed compared to $2.23 \pm 0.43$ expected from signal
processes and $0.33 \pm 0.08$ from background sources.  Four of the
muon events observed in the $e^+p$ data sample are among those first
reported and discussed in \cite{Adloff:1998aw}.  The event discussed
in \cite{origisol} is rejected from this analysis by the azimuthal
difference ($\Delta \phi_{\mu-X}$) cut. Four of the events have a
positively charged muon, three have a negative muon and in one event
the charge is not determined.

\begin{figure}[ht]
  \setlength{\unitlength}{1cm}
  \begin{picture}(12.0,15.0)
    \put(-1.0,0.0) {\epsfig{file=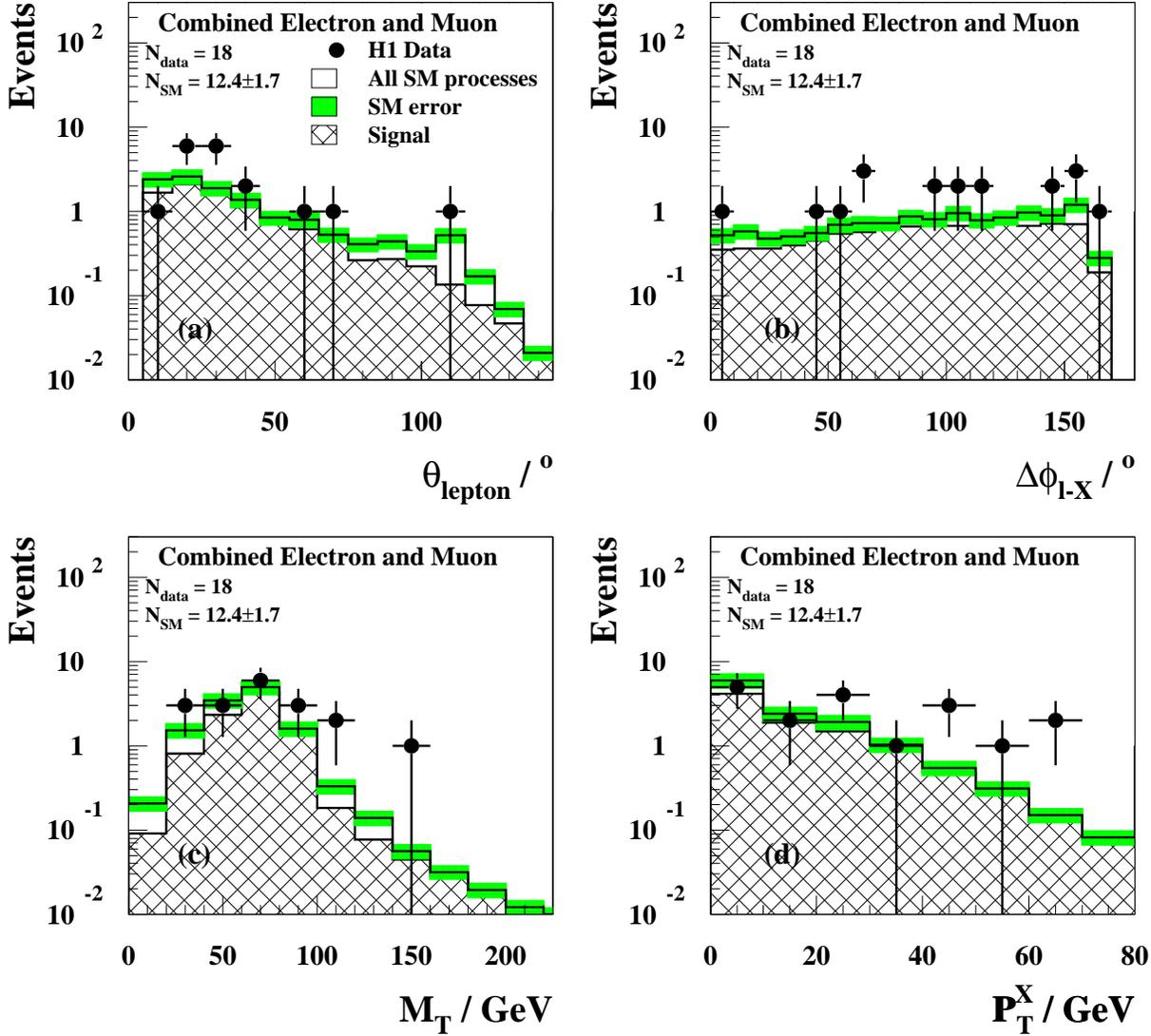,width=1.1\textwidth}}
    \put(1.45,3.35) {{\bf(c)}}
    \put(9.65,3.35) {{\bf(d)}}
    \put(1.45,10.7) {{\bf(a)}}
    \put(9.65,10.7) {{\bf(b)}}
  \end{picture}
  \caption{The final $e^+p$ data selection in the electron 
    and muon channels combined compared with the SM expectation (open
    histogram).  The total error on the SM expectation is given by the
    shaded band.  The ``signal'' component of the SM expectation is
    given by the hatched histogram. $\rm N_{data}$ is the total number
    of data events observed for each sample. $\rm N_{SM}$ is the total
    SM expectation.}
  \label{fig:emfinal} 
\end{figure}

Distributions of the selected events in lepton polar angle, azimuthal
difference, transverse mass and $P_T^X$ are shown in
figure~\ref{fig:emfinal}. The lepton--neutrino transverse mass is
defined as

\begin{equation}
  M_T  = \sqrt{(P_T^{\rm miss} + P_T^l)^2 - {(\vec{P}_T^{\rm miss} + \vec{P}_T^l)}^2},
\end{equation}
where $\vec{P}_T^{\rm miss}$ and $\vec{P}_T^l$ are the vectors of the
missing transverse momentum and isolated lepton respectively. The
figure shows the electron and muon channels combined. Also included is
the expectation of the Standard Model. The events generally have low
values of lepton polar angle and are consistent with a flat
distribution in azimuthal difference $\Delta \phi_{l-X}$, in
agreement with the expectation.  The distribution of the events in
$M_T$ is compatible with the Jacobian peak expected from $W$
production. The kinematics of the events with $P_T^X>25$ GeV are
detailed in table~\ref{evkin}.

In three of the eighteen events a further electron is detected in the
main detector ($\theta_e<176^\circ$). Taking this to be the scattered
electron and assuming that there is only one neutrino in the final
state and there is no initial state QED radiation, the
lepton--neutrino mass $M_{l\nu}$ can be reconstructed. All three
events yield masses that are consistent with the $W$ mass, having
values of $86^{+7}_{-9}$, $73^{+7}_{-7}$ and $79^{+12}_{-12}$ GeV. The
observation of a second electron in these three events is compatible
with the expectation from SM $W$ production, where approximately
$25\%$ of events have a scattered electron in the acceptance range of
the main detector.

Details of the event yields from the $e^+p$ data sample as a function
of the transverse momentum of the hadronic final state, $P_{T}^{X}$,
are given in table~\ref{etab1} and \ref{mtab1} for the electron and
muon channels respectively. The combined results for the electron and
muon channels are given in table~\ref{emtab}.  At $P_{T}^{X}<25$ GeV
eight events are seen, in agreement with the expectation from the
Standard Model.  At $P_{T}^{X}>25$ GeV ten events are seen, six of
which have $P_{T}^{X}>40$ GeV, where the signal expectation is very
low. The probability for the SM expectation to fluctuate to the
observed number of events or more is 0.10 for the full $P_{T}^{X}$
range, 0.0015 for $P_{T}^{X} > 25$ GeV and 0.0012 for $P_{T}^{X} > 40$ GeV.
The uncertainties on the SM predictions are taken into account in
calculating these probabilities.

An excess is observed at $P_{T}^{X}>25$ GeV in both sets of $e^+p$
data. In the 1994--1997 data $4$~events are observed compared to an
expectation of $0.80 \pm 0.14$. In the 1999--2000 data $6$ events are
observed compared to an expectation of $2.12 \pm 0.36$.

The method published in \cite{Adloff:1998aw} has been applied to the
1999--2000 data sample. Using this method an excess of events is also
seen at $P_{T}^{X}>25$ GeV in this new data sample: $5$~events are
observed for $2.34 \pm 0.29$ expected.  These $5$ events selected by
the method of the previously published analysis are also found by the
analysis presented in this paper.

\subsection{Cross Section}
\label{sec:crosssec}

The observed number of events in the $e^+p$ data sample is corrected
for acceptance and detector effects to obtain a cross section for all
processes yielding genuine isolated electrons or muons and missing
transverse momentum.  This is defined for the kinematic region
$5^\circ < \theta_l < 140^\circ$, $P_T^l > 10$ GeV, $P_T^{\rm
  miss}>12$ GeV and $D_{jet} > 1.0$ at a centre of mass
energy\footnote{Assuming a linear dependence of the cross section on
  the proton beam energy.} of $\sqrt{s}=312$~GeV.  The definition of
isolated electrons or muons includes those from leptonic tau decay.
The generator EPVEC is used to calculate the detector acceptance $A$
for this region of phase space. The acceptance accounts for trigger and
detection efficiencies and migrations. The cross section is thus
\begin{equation}
  \label{eq:isolepxs}
  \sigma = \frac{({\rm N_{data}} - {\rm N_{bgd}})}{{\cal L}{A}},
\end{equation}
where ${\rm N_{data}}$ is the number of events observed, ${\rm
  N_{bgd}}$ is the number of events expected from
processes treated here as background (see section \ref{sec:monte}) and
${\cal L}$ is the integrated luminosity of the data sample.

The cross section integrated over the full $P_T^X$ range is
\begin{equation}
  \sigma = 0.31 \pm 0.10 \pm 0.04 {\rm~pb},
\end{equation}
where the first error is statistical and the second is systematic
(calculated as described in section~\ref{sec:expsyst}). 

This result is compatible with the SM signal expectation of $0.237 \pm
0.036$~pb, dominated by the process $ep \rightarrow eWX$, calculated
at NLO \cite{Diener:2002if,spiraprivate}.  The small signal components
from $ep \rightarrow \nu WX$ and $Z$ production are calculated with
EPVEC \cite{Baur:1991pp} as explained in section~\ref{sec:monte}. The
cross section is presented in table~\ref{cross_section} split into the
regions $P_T^X < 25$~GeV and $P_T^X > 25$~GeV. Whilst the cross
section in the low $P_T^X$ region agrees within errors with the
prediction, in the high $P_T^X$ region it exceeds the expectation.
Table \ref{cross_section} also includes two signal calculations in
which all components are calculated at LO
\cite{Diener:2002if,Baur:1991pp}. The calculation in
\cite{Baur:1991pp} is the default calculation implemented in the event
generator EPVEC. All the calculations agree within the uncertainties.

\section{Search for {\boldmath $W$} Production in the Hadronic Decay Channel}
\label{sec:hadchan}

Since the dominant SM process that produces events with isolated
charged leptons and missing transverse momentum is $W$ production, it
is interesting to search for $W$ bosons decaying hadronically. The
search for hadronic $W$ decays is performed using events with two high
transverse momentum jets in $117.3$ pb$^{-1}$ of $e^+p$ and $e^-p$
data from the period 1995--2000.

Events are selected with at least two hadronic jets, reconstructed
using an inclusive $k_T$ algorithm, with a transverse momentum $P_T$
greater than $25$ GeV for the leading jet and greater than $20$ GeV
for the second highest $P_T$ jet. The minimum $P_T$ of any further jet
considered in the event is set to $5$ GeV.  The pseudorapidity $\eta$
of each jet is restricted to the range $-0.5<\eta<2.5$. The dijet
combination with invariant mass $M_{jj}$ closest to the $W$ mass is
selected as the $W$ candidate. The resolution of the reconstructed $W$
mass is approximately $5$~GeV. $P_T^X$ is defined as the transverse
momentum of the hadronic system after excluding the $W$ candidate
jets.
 
A cut on the missing transverse momentum $P_T^{\rm miss}<20$ GeV is
applied to reject CC events and non--$ep$ scattering background. NC
events where the electron is misidentified as a jet are rejected
\cite{gillesthesis,saschathesis}.  The final selection is made with
the cuts $M_{jj}>70$ GeV and $|\mbox{cos}\hat{\theta}|<0.6$, where
$\hat{\theta}$ is the decay angle in the $W$ rest frame, with the $W$
flight direction in the laboratory frame taken as the quantisation
axis.  This phase space is chosen to optimise the acceptance for $W$
events and reduce other SM contributions. The overall selection
efficiency for SM $W$ production is 43\% and is 29\% for
$P_T^X>40$~GeV.

The main physics background to this search is the 
production of jets via hard partonic scattering, which is modelled by 
PYTHIA and RAPGAP for the photoproduction and deep inelastic regimes
respectively. 
The predicted cross section is increased by a factor of $1.2$ in
order to match the observed number of events outside the signal region.

The systematic uncertainty on the background prediction includes
parton distribution function uncertainties, the uncertainty on the jet
energy scale and uncertainties due to the misidentification of an
electron as a jet.  In quadratic sum these give a total systematic
error on the background prediction of
$23\%$~\cite{gillesthesis,Adloff:2002au}. The SM $W$ production rate
has a theoretical error of $15\%$, which is added in quadrature to the
experimental uncertainties, resulting in an overall error of $21\%$.

\begin{figure}[ht]
  \setlength{\unitlength}{1cm}
  \begin{picture}(12.0,4.0)
    \put(-1.0,0.0) {\epsfig{file=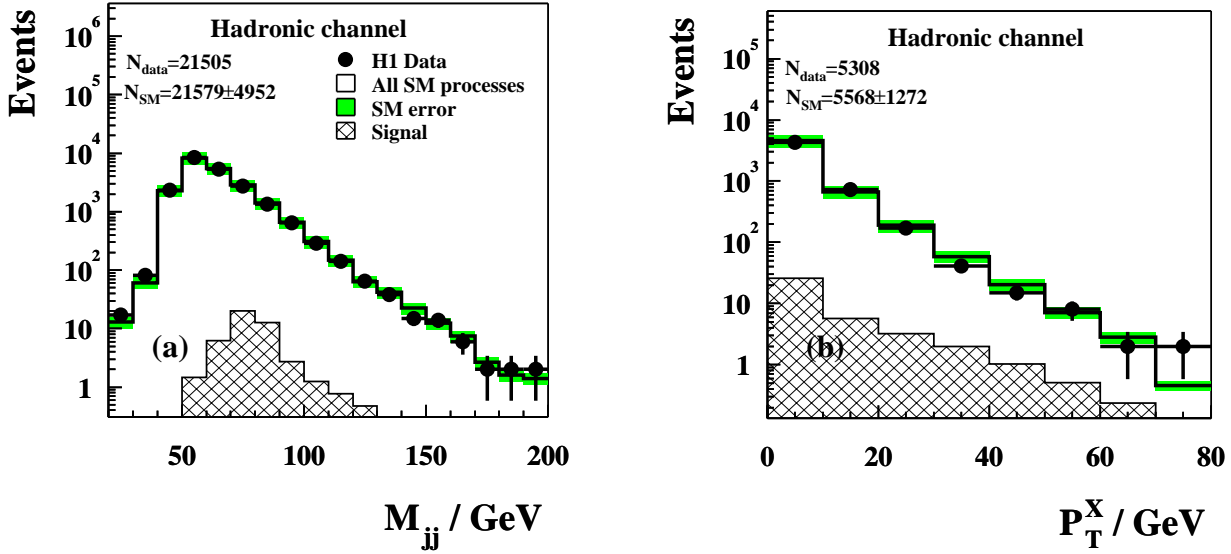,width=1.1\textwidth}}
    \put(1.5,2.75) {{\bf(a)}}
    \put(10.2,2.75) {{\bf(b)}}
  \end{picture}
  \caption{
    The dijet mass distribution $M_{jj}$ (a) and the $P_T^X$
    distribution for $M_{jj} > $ 70 GeV (b) compared with the SM
    expectation (open histogram) in the $W$ hadronic decay channel
    search. The total error on the SM expectation is given by the
    shaded band. The $W$ production component of the SM expectation is
    given by the hatched histogram.  $\rm N_{data}$ is the total number
    of data events observed for each sample. $\rm N_{SM}$ is the total
    SM expectation.}
\label{fig:whad}
\end{figure}

The $M_{jj}$ distribution (without the $M_{jj}$ cut) and the $P_T^X$
distribution (with all cuts) of the selected data are compared to the
Standard Model in figure~\ref{fig:whad}.  The final data selected show
overall agreement with the SM expectation up to the highest $P_T^X$
values. At $P_T^X>25$~GeV, $126$ events are observed compared to $162
\pm 36$ expected with $5.3 \pm 1.1$ from $W$ production. The
expectation is dominated by QCD multi--jet production.  For
$P_T^X>40$~GeV $27$ events are observed in the data, compatible with
the expectation of $30.9 \pm 6.7$, where the $W$ contribution amounts
to $1.9 \pm 0.4$ events.  Although there is increasing sensitivity to
$W$ production with increasing $P_T^X$, it is at present not possible
to conclude from the hadronic channel whether the observed excess of
events with an isolated electron or muon with missing transverse
momentum at high $P_T^X$ is due to $W$ production.

\section{Summary}
\label{sec:summary}

A search for events with isolated electrons or muons and missing
transverse momentum has been performed in $e^+p$ and $e^-p$ data,
using the complete HERA I (1994-2000) data sample.  The selection has
been optimised to increase the acceptance for $W$ production events
and it extends to lower values of hadronic transverse momentum
$P_{T}^{X}$ than in previous publications.

One electron event and no muon events are observed in the $e^-p$ data,
consistent with the expectations of $1.69 \pm 0.22$ and $0.37 \pm
0.06$ for the electron and muon channels respectively in this
relatively low luminosity data sample.  In the $e^+p$ data sample $10$
events are observed in the electron channel and $8$ in the muon
channel.  These events are kinematically consistent with $W$
production. The expected numbers of events from the Standard Model are $9.9 \pm
1.3$ and $2.55 \pm 0.44$ for the electron and muon channels
respectively.

At low $P_T^X$, the number of observed events in both channels
is consistent with the expectation. At $P_T^X> 25$ GeV, however, the
$10$ observed events exceed the SM prediction of $2.92 \pm 0.49$. An
excess of events is observed in both the 1994--1997 and the 1999--2000
$e^+p$ data samples. The observed events are used to make a
measurement of the cross section for all processes producing isolated
electrons or muons and missing transverse momentum in the kinematic
region studied.

In a separate search for hadronic $W$ decays, agreement with the
SM expectation is found up to the highest $P_T^X$ values.
The high background in this channel, however, does not allow one to
conclude whether the excess of isolated leptons with missing $P_T$ at
high $P_T^X$ is due to $W$ production.

\section*{Acknowledgements}

We are grateful to the HERA machine group whose outstanding efforts have
made this experiment possible.  We thank the engineers and technicians
for their work in constructing and maintaining the H1 detector, our
funding agencies for financial support, the DESY technical staff for
continual assistance and the DESY directorate for support and
for the hospitality
which they extend to the non--DESY members of the collaboration. The
authors wish to thank K.P.~Diener and M.~Spira for many useful
discussions and for providing the NLO calculation for $W$ production.



\begin{table}[htb]
  \begin{center}
    \renewcommand{\arraystretch}{1.2}
    \begin{tabular}{|lll||l|l|l|l|l|}
      \hline
      Run & Event & Lepton  & $P_T^{l}$ /GeV         & $P_T^{X}$ /GeV       & $M_T$ /GeV      & $M_{l \nu}$ /GeV & Charge \\ 
      \hline \hline
      236176 & 3849 & $e$& $10.1^{+0.4}_{-0.4}$   & $25.4^{+2.8}_{-2.5}$ 
      & $26.1^{+1.1}_{-1.1}$ & & unmeasured \\ \hline \hline
      186729 & 702  &$\mu^+$  & $51^{+11}_{-17}$ & $66.7^{+4.9}_{-4.9}$
      & $43^{+13}_{-22}$ & &  $+$ (4.0$\sigma$)\\ \hline
      188108 & 5066 &$\mu^-$& $41.0^{+4.3}_{-5.5}$   & $26.9^{+2.2}_{-2.3}$ 
      & $81.3^{+8.2}_{-11}$  & $86.1^{+6.8}_{-8.7}$ & $-$ (8.3$\sigma$)\\ \hline
      192227 & 6208 &$\mu^-$  & $73^{+9}_{-12}$  & $60.5^{+5.5}_{-5.4}$ 
      & $74^{+20}_{-25}$ & & $-$  (7.0$\sigma$)  \\ \hline
      195308 & 16793 & $\mu^+$& $60^{+12}_{-19}$ & $33.3^{+3.6}_{-3.6}$ 
      & $85^{+25}_{-37}$ &  & $+$ (4.2$\sigma$)  \\ \hline  
      248207 & 32134 & $e^+$& $32.0^{+0.8}_{-0.9}$   & $42.7^{+3.9}_{-4.1}$
      &   $62.8^{+1.8}_{-1.8}$  &  &   $+$ (15$\sigma$) \\ \hline        
      252020 & 30485 & $e^+$& $25.3^{+1.0}_{-1.0}$   & $44.3^{+3.6}_{-3.6}$ 
      & $50.6^{+1.9}_{-2.0}$   & $79^{+12}_{-12}$ &   $+$ (40$\sigma$) \\ \hline        
      266336 & 4126 &$\mu^+$&  $19.7^{+0.7}_{-0.8}$  & $51.5^{+3.8}_{-4.0}$ 
      & $69.2^{+2.4}_{-2.6}$   & & $+$ (26$\sigma$)\\ \hline
      268338 & 70014 &$e^+$ &  $32.1^{+0.9}_{-0.8}$  & $46.6^{+3.3}_{-3.3}$ 
      & $87.7^{+2.5}_{-2.4}$   & &$+$  (5.1$\sigma$)  \\ \hline
      270132 & 73115 &$\mu$ & $64^{+38}_{-55}$ & $27.3^{+3.9}_{-3.9}$ 
      & $140^{+71}_{-83}$ & & $-$ (0.6$\sigma$) \\ \hline
      275991 & 29613 & $e^+$& $37.7^{+1.0}_{-1.1}$   & $28.4^{+5.7}_{-5.9}$ 
      & $74.7^{+2.3}_{-2.4}$ & &$+$ (37$\sigma$) \\ \hline
    \end{tabular}
    \caption{Kinematics and lepton charges of the events at high $P_T^X$
      ($>25$ GeV).  The invariant mass $M_{l \nu}$ is only calculated
      for those events with an observed scattered electron. The
      significance of the charge measurement in numbers of standard
      deviations is given in brackets after the sign. The first event
      listed was observed in $e^-p$ data.  The rest were observed in
      $e^+p$ data.}
    \label{evkin}
  \end{center}
\end{table}

\begin{table}[htb]
  \renewcommand{\arraystretch}{1.1}
  \begin{center}
    \begin{tabular}{|r||c|c||c|c|} \hline
      Electron & H1 Data & SM expectation & SM Signal & Other SM processes\\ \hline
      \hline  
      $P_T^X<12$~GeV         & 5 & 6.40 $\pm$ 0.79  & 4.45 $\pm$ 0.70  & 1.95 $\pm$ 0.36 \\
      \hline  
      $12<P_{T}^{X}<25$~GeV & 1 & 1.96 $\pm$ 0.27  & 1.45 $\pm$ 0.24  & 0.51 $\pm$ 0.12  \\
      \hline
      $25<P_{T}^{X}<40$~GeV & 1 & 0.95 $\pm$ 0.14  & 0.82 $\pm$ 0.13  & 0.13 $\pm$ 0.04  \\
      \hline
      $P_{T}^{X}>40$~GeV & 3 & 0.54 $\pm$ 0.11  & 0.45 $\pm$ 0.11  & 0.09 $\pm$ 0.04  \\
      \hline
    \end{tabular}
  \end{center}
  \caption{Observed and predicted numbers of events in the electron channel for all $e^+p$ data.}
  \label{etab1}
\end{table}

\begin{table}[htb]   
  \renewcommand{\arraystretch}{1.1}
  \begin{center}
    \begin{tabular}{|r||c|c||c|c|} \hline
      Muon & H1 Data & SM expectation & SM Signal & Other SM processes\\ \hline
      \hline  
      $12<P_{T}^{X}<25$~GeV & 2 & 1.11 $\pm$ 0.19  & 0.94 $\pm$ 0.18  & 0.17 $\pm$ 0.05  \\
      \hline
      $25<P_{T}^{X}<40$~GeV & 3 & 0.89 $\pm$ 0.14  & 0.77 $\pm$ 0.14  & 0.12 $\pm$ 0.03  \\
      \hline
      $P_{T}^{X}>40$~GeV & 3 & 0.55 $\pm$ 0.12  & 0.51 $\pm$ 0.12  & 0.04 $\pm$ 0.01  \\
      \hline
    \end{tabular}
  \end{center}
  \caption{Observed and predicted numbers of events in the muon channel for all $e^+p$ data.}
  \label{mtab1}        
\end{table}

\begin{table}[htb]
  \renewcommand{\arraystretch}{1.1}
  \begin{center}
    \begin{tabular}{|r||c|c||c|c|} \hline
      Electron and Muon & H1 Data & SM expectation & SM Signal & Other SM processes\\ \hline
      \hline  
      $P_{T}^{X}<12$~GeV   &  5  & 6.40 $\pm$ 0.79  & 4.45 $\pm$ 0.70  & 1.95 $\pm$ 0.36 \\
      \hline  
      $12<P_{T}^{X}<25$~GeV & 3  & 3.08 $\pm$ 0.43  & 2.40 $\pm$ 0.40  & 0.68 $\pm$ 0.14  \\
      \hline
      $25<P_{T}^{X}<40$~GeV & 4  & 1.83 $\pm$ 0.27  & 1.59 $\pm$ 0.26  & 0.24 $\pm$ 0.06  \\
      \hline
      $P_{T}^{X}>40$~GeV    & 6  & 1.08 $\pm$ 0.22  & 0.96 $\pm$ 0.22  & 0.12 $\pm$ 0.04  \\
      \hline
    \end{tabular}
  \end{center}
  \caption{Observed and predicted numbers of events in the electron and muon channels combined for all $e^+p$ data. Only the electron channel contributes for
    $P_{T}^{X}<12$ GeV.}
  \label{emtab}        
\end{table}

\begin{table}[htb]
  \renewcommand{\arraystretch}{1.1}
  \begin{center}
    \begin{tabular}{|c||c|c||c|c|} \hline
      &  \multicolumn{4}{c|}{Cross Section  / pb}\\ \hline
      & Measured & SM NLO & SM LO  & SM LO \\
      & & & Diener {\it et al.} & Baur {\it et al.} \\ \hline
      \hline
      $P_T^X<25$~GeV    &  0.146 $\pm$ 0.081 $\pm$ 0.022  &  0.194 $\pm$ 0.029 & 0.147 $\pm$ 0.044 & 0.197 $\pm$ 0.059 \\ \hline
      $P_T^X>25$~GeV    &  0.164 $\pm$ 0.054 $\pm$ 0.023  &  0.043 $\pm$ 0.007 & 0.041 $\pm$ 0.012 & 0.049 $\pm$ 0.015 \\ \hline
    \end{tabular}
  \end{center}
  \caption{The measured cross section for events with an isolated high energy 
    electron or muon with missing transverse momentum. The cross sections are 
    calculated in the kinematic region: $5^\circ < \theta_l < 140^\circ$; 
    $P_T^l > 10$~GeV; $P_T^{\rm miss} > 12$~GeV and $D_{jet} > 1.0$. Also shown 
    are the signal expectations from the Standard Model where the dominant 
    contribution $ep \rightarrow eWX$ is calculated at next to leading order (SM NLO) \cite{Diener:2002if,spiraprivate}
    and at leading order (SM LO) \cite{Diener:2002if} and \cite{Baur:1991pp}.}
  \label{cross_section}
\end{table}

\end{document}